\documentclass{webofc}
\usepackage[varg]{txfonts}   
\begin{document}
\title{Exploration of hadronization through heavy flavor production at the future Electron-Ion Collider}
%
%

\author{\firstname{Xuan} \lastname{Li}\inst{1}\fnsep\thanks{\email{xuanli@lanl.gov}} 
}

\institute{Physics division, Los Alamos National Laboratory, Los Alamos, USA 
          }

\abstract{%
The future Electron-Ion Collider will utilize high-luminosity high-energy electron+proton ($e+p$) and electron+nucleus ($e+A$) collisions to solve several fundamental questions in the high energy nuclear physics field. Heavy flavor products play an important role in constraining the initial-state nucleon/nucleus parton distribution functions especially in the high and low Bjorken-x ($x_{BJ}$) region and exploring the final-state parton propagation and hadronization processes under different nuclear medium conditions. Latest simulation studies of heavy flavor hadron and jet measurements with the EIC project detector conceptual design will be discussed. The projected statistical accuracy of heavy flavor jet and heavy flavor hadron inside jet measurements in comparison with latest theoretical calculations will be presented.
 }
\maketitle
\section{Introduction}
\label{intro}
The future Electron-Ion Collider (EIC) to be built at Brookhaven National Laboratory will start construction in 2025. A series of high-luminosity electron+proton ($e+p$) and electron+nucleus ($e+A$) collisions at the center of mass energies of 29-141~GeV will be operated at the EIC \cite{eic_YR} to explore the inner structure of nucleon/nucleus, probe the parton density extreme and study how matter is formed from quarks and gluons. Latest conceptual design of the EIC project detector led by the ePIC collaboration can achieve good momentum and primary/decay vertex resolution, precise particle identification and good energy determination for hadron and jet reconstruction. Heavy quarks are produced early in particle collisions and remain their flavors until the final-state hadrons are formed. In addition to constraining the initial-state nucleon/nucleus parton distribution functions, heavy flavor productions in $e+p$ and $e+A$ collisions are ideal probes to explore the final-state parton propagation and hadronization process under different nuclear medium conditions. Progresses of latest simulation studies of heavy flavor di-jet and heavy flavor hadron inside jet measurements with the current ePIC detector conceptual design will be discussed in the following sections.

\section{Heavy flavor hadron and jet reconstruction in simulation}
\label{sec-1}
\begin{figure}[ht]
\centering
\includegraphics[width=0.27\textwidth,clip]{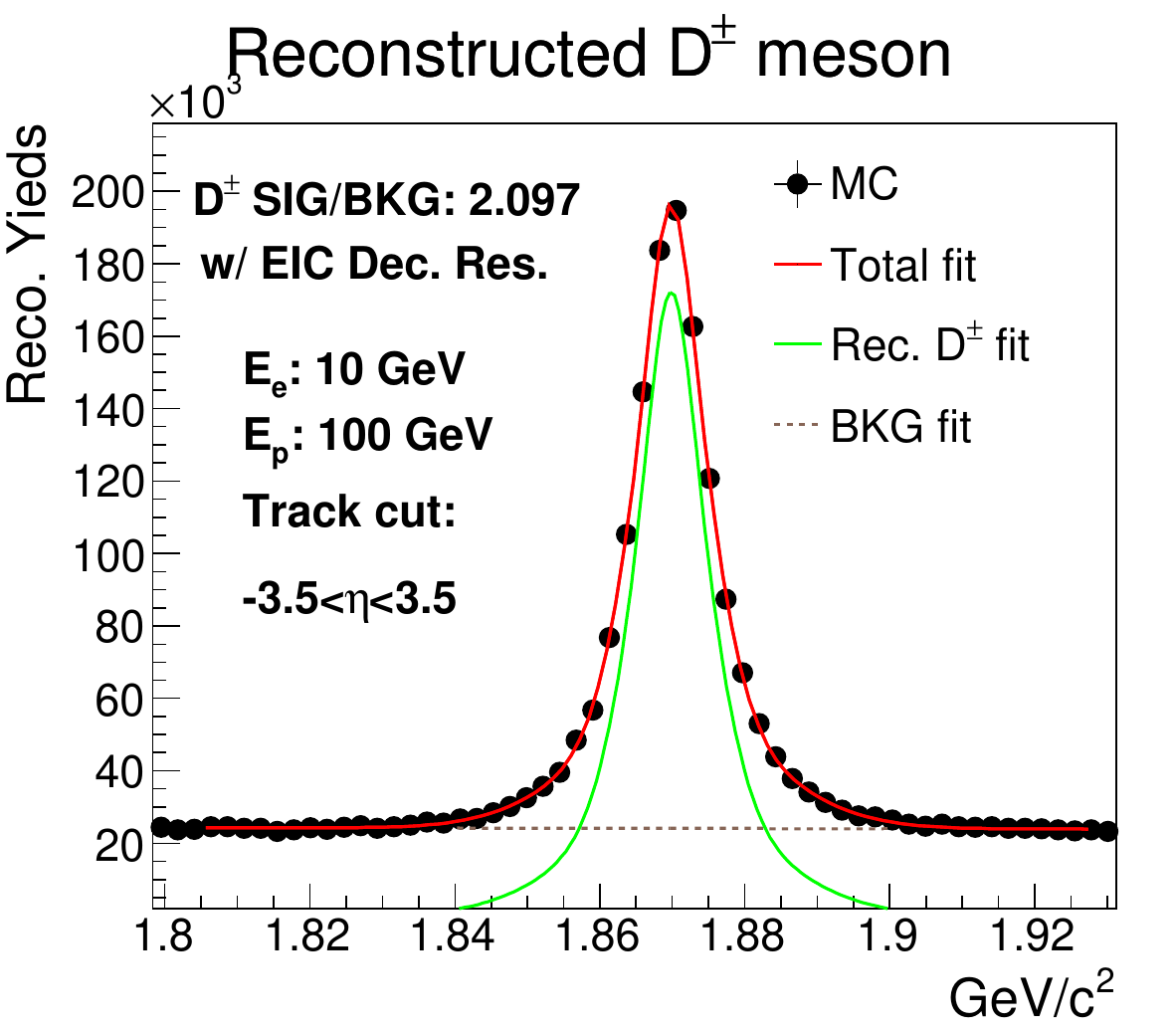}
\includegraphics[width=0.27\textwidth,clip]{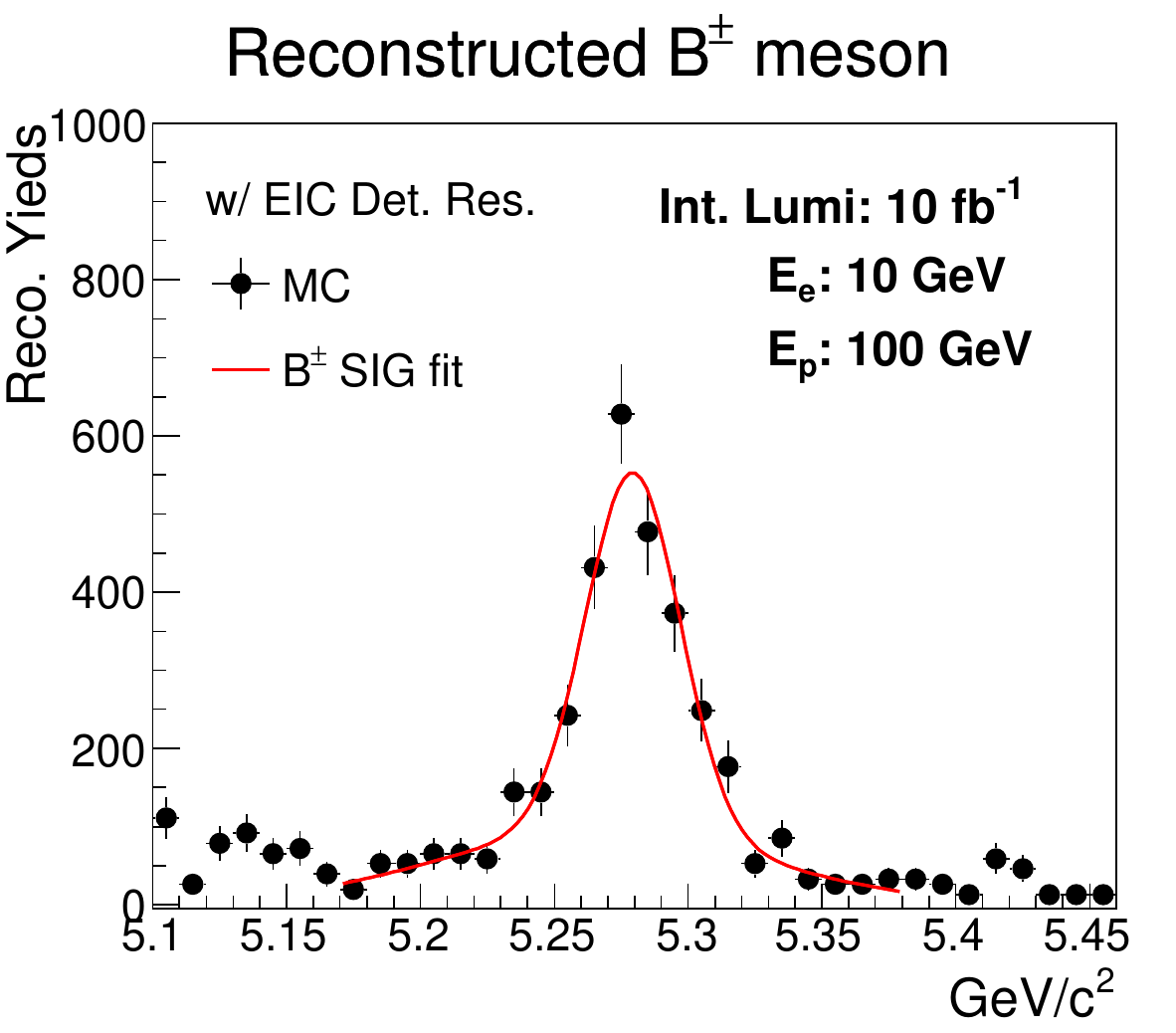}
\includegraphics[width=0.27\textwidth,clip]{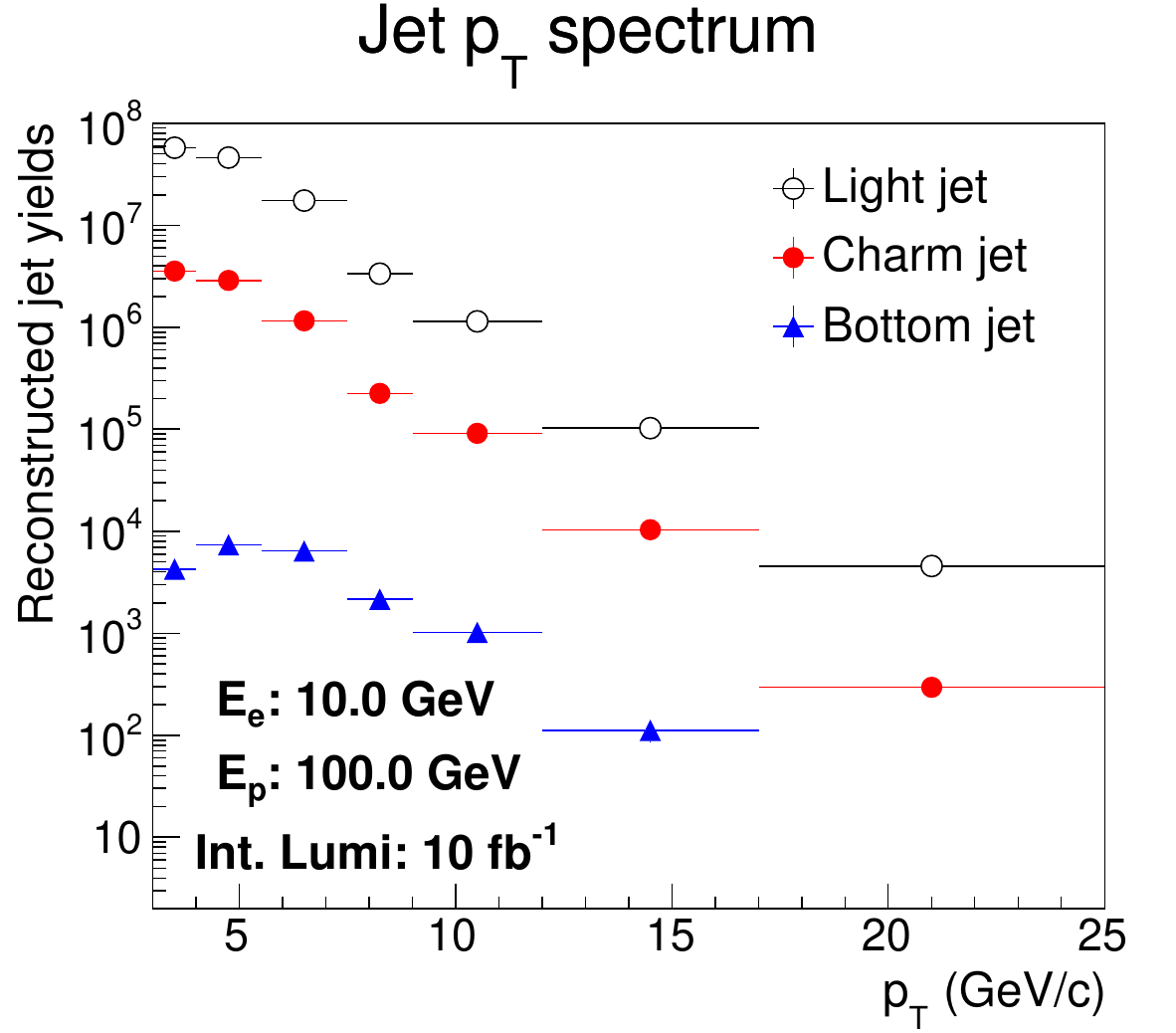}
\caption{Invariant mass spectrums of reconstructed $D^{\pm}$ (left) and reconstructed $B^{\pm}$ (right) in 63.2~GeV $e+p$ simulation with the current ePIC detector performance. The transverse momentum ($p_{T}$) spectrums of reconstructed jets with light (black open circles), charm (red closed circles) and bottom (blue closed triangles) flavors in the same simulation sample are shown in the right panel.}
\label{fig-1}       
\end{figure}

A standalone simulation framework \cite{lanl_hf1}, which includes the event generation in PYTHIA 8 \cite{py8}, a particle smearing package with the ePIC detector performance evaluated in GEANT4 \cite{geant4} and beam remnant background, has been used to reconstruct heavy flavor hadrons and jets in $e+p$ collisions. Through several selection cuts such as matching transverse Distance of Closest Approach ($DCA_{2D}$) of heavy flavor decay daughter candidates, the reconstruction capability of charm/bottom hadrons and jets at the EIC has been validated. The left and middle panels of Figure~\ref{fig-1} show the mass spectrums of reconstructed $D^{\pm}$s ($D^{\pm} \rightarrow K^{\mp}\pi^{\pm}\pi^{\pm}$) and reconstructed $B^{\pm}$s ($B^{\pm} \rightarrow J/\psi (\rightarrow l^{+}l^{-}) + K^{\pm}$) in simulation using the current ePIC detector performance in 10 GeV electron and 100 GeV proton collisions. The transverse momentum ($p_{T}$) spectrums of reconstructed jets with light (u,d,s,g), charm and bottom flavors in the same $e+p$ simulation are shown in the right panel of Figure~\ref{fig-1}. Jets are reconstructed with the anti-$k_{T}$ algorithm and the jet cone radius is selected at 1.0. The jet flavor is tagged with the associated hadron flavor with the most displaced secondary vertex found inside the jet. Clear and pronounced signals of charm and bottom hadrons/jets can be achieved with around one-year $e+p$ operation at the EIC, which is equivalent to 10~$fb^{-1}$ integrated luminosity.

\begin{figure}[ht]
\centering
\includegraphics[width=0.32\textwidth,clip]{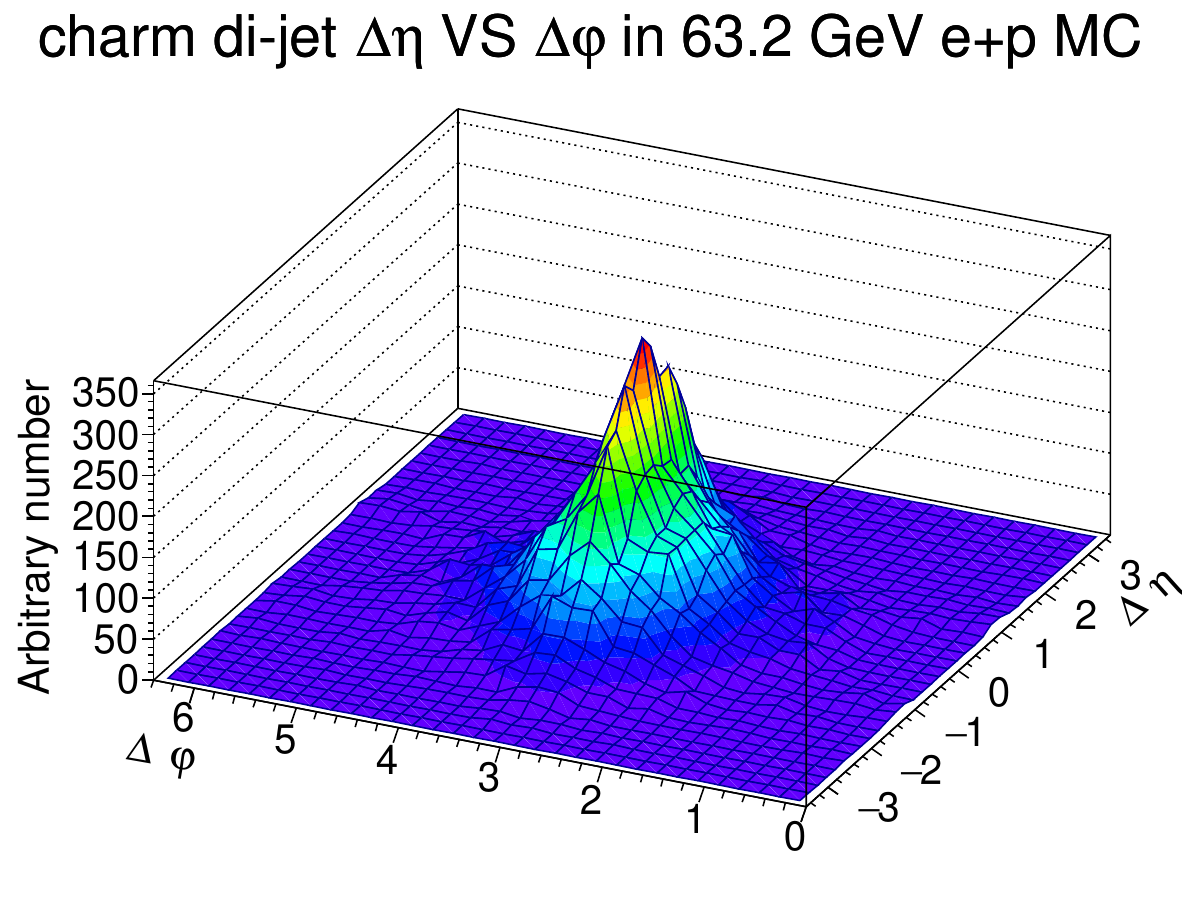}
\includegraphics[width=0.32\textwidth,clip]{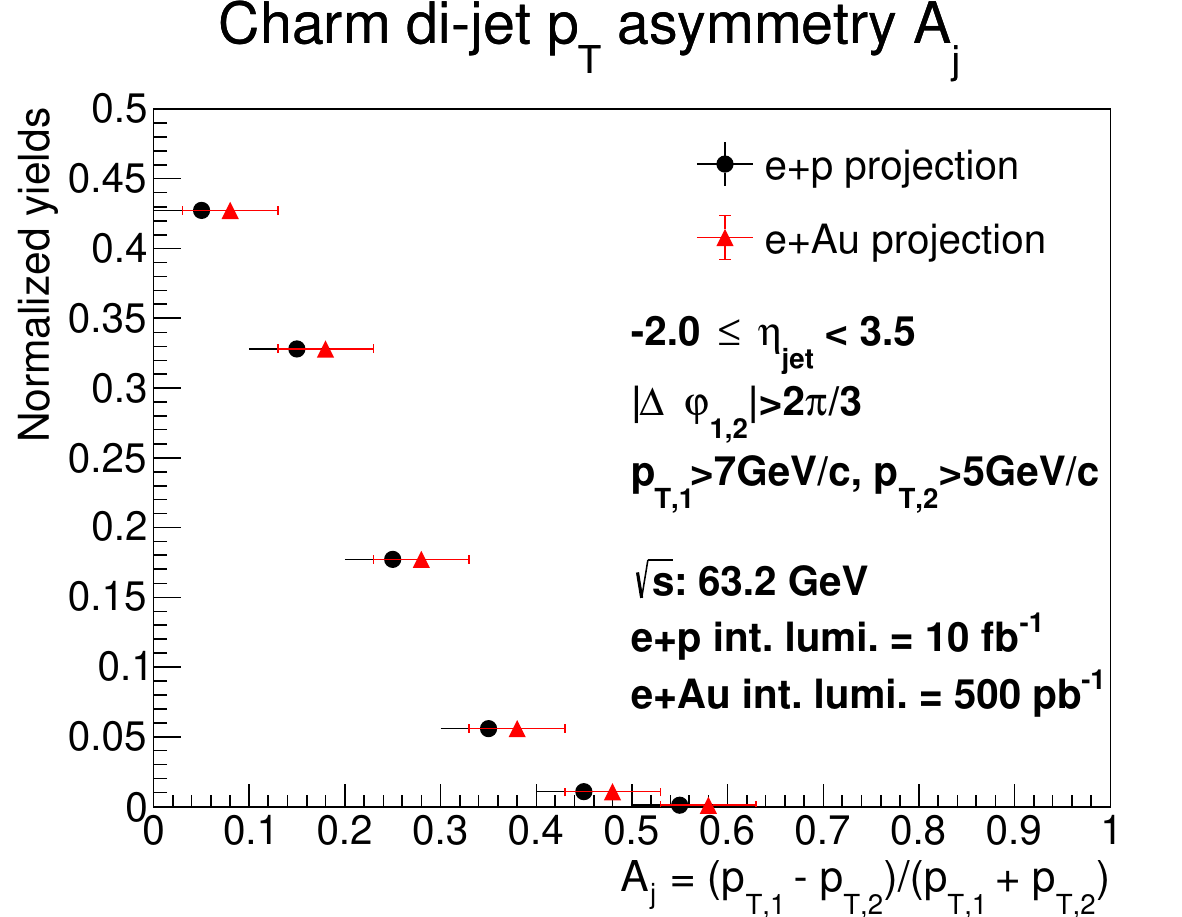}
\caption{Left: Pseudorapidity difference $\Delta\eta$ VS azimuthal angle difference $\Delta\varphi$ of charm di-jets in 63.2 GeV $e+p$ simulation. Right: Normalized yields of charm di-jet transverse momentum asymmetry $A_{j} = (p_{T,1} - p_{T,2})/(p_{T,1} + p_{T,2})$ in 63.2 GeV $e+p$ (black) and $e+Au$ (red) collisions, which are evaluated in simulation with the current ePIC detector performance.}
\label{fig-2}       
\end{figure}

\section{Heavy flavor di-jet correlation studies for the EIC}
\label{sec-2}
Kinematics of initial-state gluons or heavy quarks can be better constrained by heavy flavor di-jet production compared to inclusive heavy flavor hadron and jet measurements. Meanwhile heavy quark energy loss mechanism and their medium transport properties can be systematically studied in $e+p$ and $e+A$ collisions at the EIC. The probability of finding charm di-jets has been first evaluated in $e+p$ simulation through the same jet tagging method discussed in  Sect.~\ref{sec-1}. The left panel of Figure~\ref{fig-2} presents the charm di-jet correlations in the pseudorapidity $\eta$ and azimuthal angle $\varphi$ phase space in 63.2 GeV $e+p$ simulation. A minimum jet $p_{T}$ cut ($p_{T} > 5$ GeV/c) is applied to charm jet reconstruction to reduce the underlying event contribution. Through requiring the leading charm jet $p_{T} > 7$ GeV/c; the sub-leading charm jet $p_{T} > 5$ GeV/c; reconstructed charm jet pseudorapdity within $-2.0<\eta<3.5$; and the charm di-jets in the back-to-back region ($\Delta \varphi_{1,2} > 2\pi/3$), the projected normalized yields of charm di-jet transverse momentum asymmetry $A_{j} = (p_{T,1} - p_{T,2})/(p_{T,1} + p_{T,2})$ in 63.2 GeV $e+p$ (black) and $e+Au$ (red) collisions are shown in the right panel of Figure~\ref{fig-2}. The $A_{j}$ distribution difference between $e+p$ and $e+Au$ collisions is related with the medium density difference between the two collision systems. Further studies will be carried out for different $e+A$ collisions with the mass number from 2 to 238 to systematically study the charm quark transport coefficient properties under different cold nuclear medium conditions.

\section{Heavy flavor hadron inside jet studies for the EIC}
\label{sec-3}
\begin{figure}[ht]
\centering
\includegraphics[width=0.27\textwidth,clip]{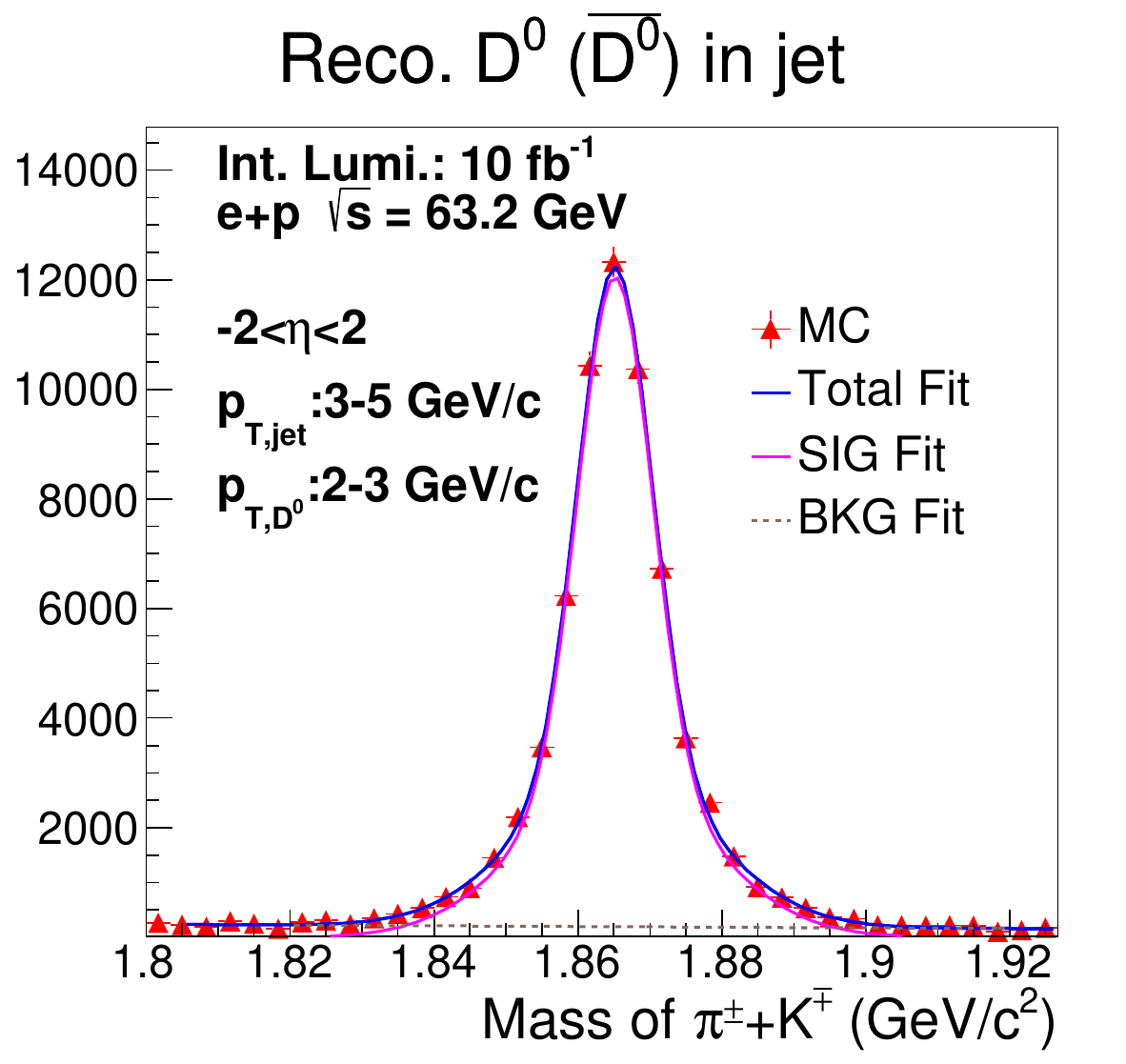}
\includegraphics[width=0.27\textwidth,clip]{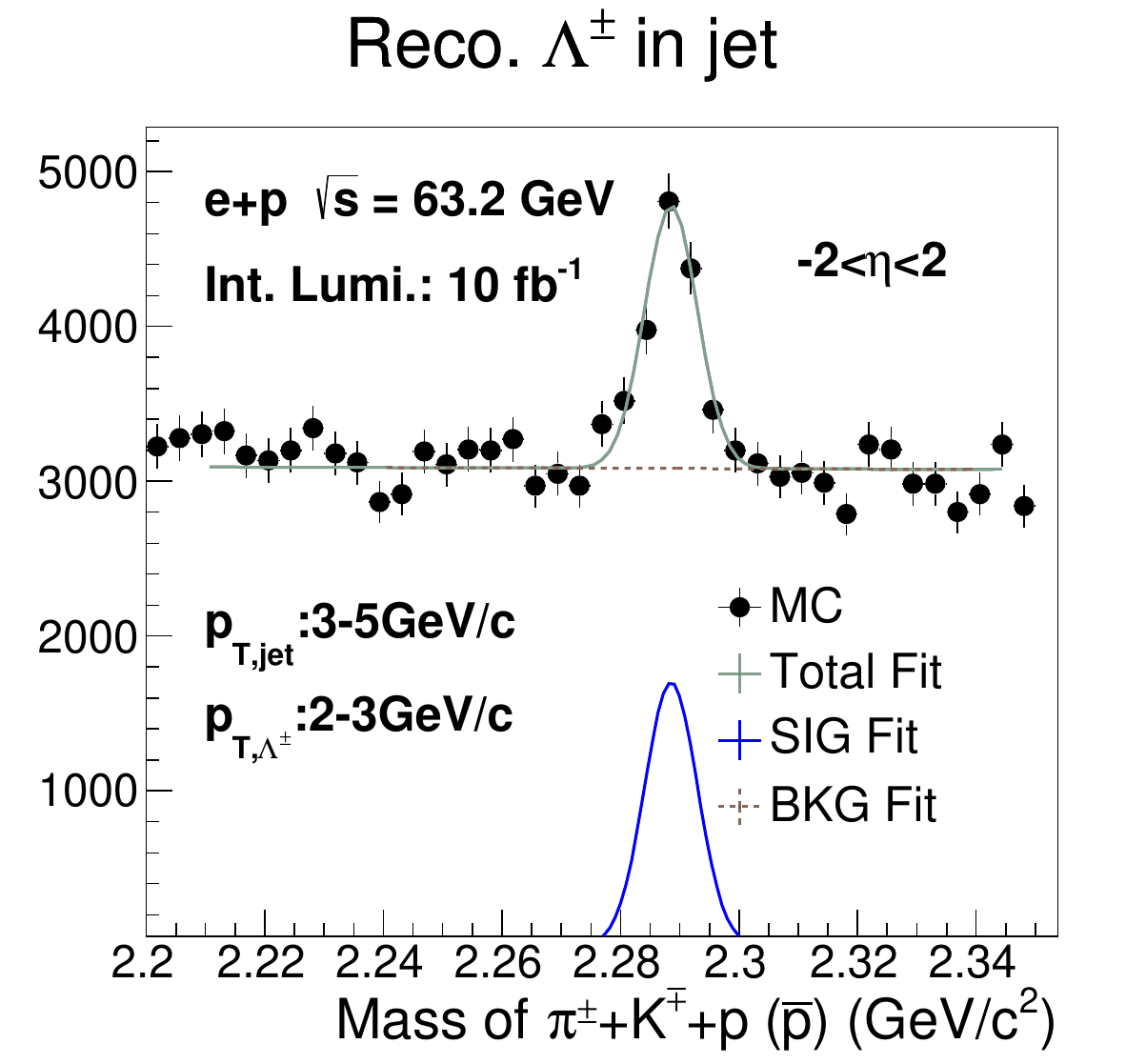}
\includegraphics[width=0.29\textwidth,clip]{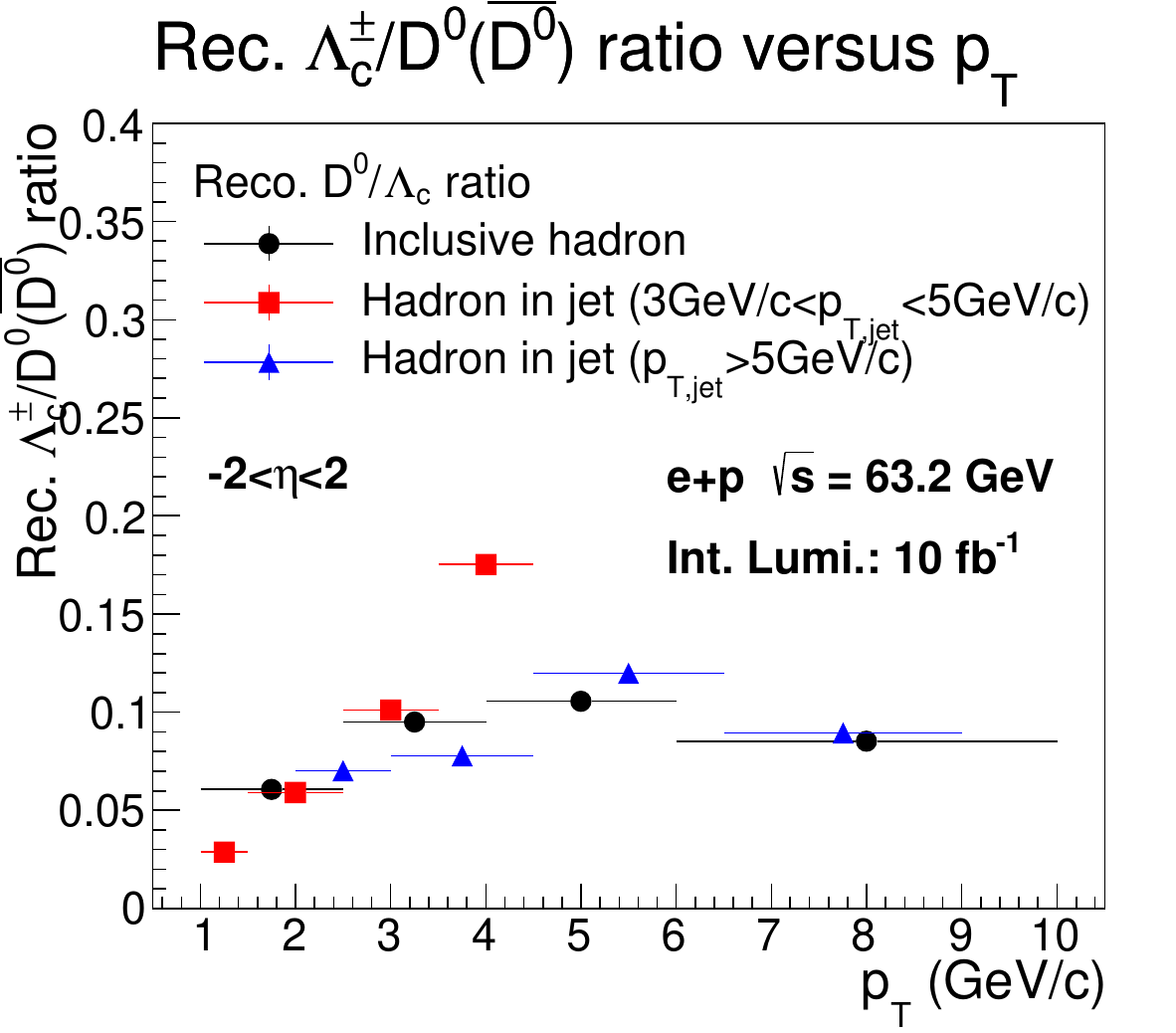}
\caption{Invariant mass spectrums of reconstructed $D^{0}$ ($\bar{D^{0}}$) inside jet (left) and reconstructed $\Lambda_{c}^{\pm}$ inside jet in simulation for 63.2 GeV $e+p$ collisions, which are evaluated with the current ePIC detector performance. Reconstructed $\Lambda_{c}$ over $D^{0}$ ratios for inclusive hadrons (black closed circles), charm hadron inside jet with 3~GeV/c$<p_{T,jet}<$5~GeV/c (red closed squares), and charm hadron inside jet with $p_{T,jet}>$5~GeV/c (blue closed triangles) are shown in the right.}
\label{fig-3}       
\end{figure}

Heavy flavor hadron inside jet measurements have better sensitivity to explore the heavy quark hadronization process. The universality of charm quark fragmentation function has been challenged by recent $\Lambda_{c}$ over $D^{0}$ ratio measurements in $p+p$ and $A+A$ collisions at Relativistic Heavy Ion Collider (RHIC) and the Large Hadron Collider (LHC). Inclusive hadron cross section contains contributions from initial-state parton kinematics and final-state parton fragmentation and hadronization processes. To isolate initial and final state effects, heavy flavor hadron inside jet productions have been studied in simulation with the ePIC detector performance. The left and middle panels of Figure~\ref{fig-3} illustrate the invariant mass distributions of reconstructed $D^{0}$ ($\bar{D^{0}}$) inside a jet and reconstructed $\Lambda_{c}^{\pm}$ inside a jet with reconstructed jet $p_{T}$ within the 3-5~GeV/c region and reconstructed $D^{0}$ ($\bar{D^{0}}$) or $\Lambda_{c}^{\pm}$ $p_{T}$ within the 2-3 GeV/c region in 63.2 GeV $e+p$ collisions. All reconstructed charm hadrons and jets are required to be within $-2<\eta<2$ pseudorapidity region. The right panel of Figure~\ref{fig-3} shows the hadron $p_{T}$ dependent projected reconstructed $\Lambda_{c}/D^{0}$ ratios in 63.2~GeV $e+p$ collisions for three different scenarios: i) inclusive charm hadrons; ii) charm hadron inside jet with 3~GeV/c$<p_{T,jet}<$5~GeV/c; and iii) charm hadron inside jet with $p_{T,jet}>$5~GeV/c.

\begin{figure}[ht]
\centering
\includegraphics[width=0.8\textwidth,clip]{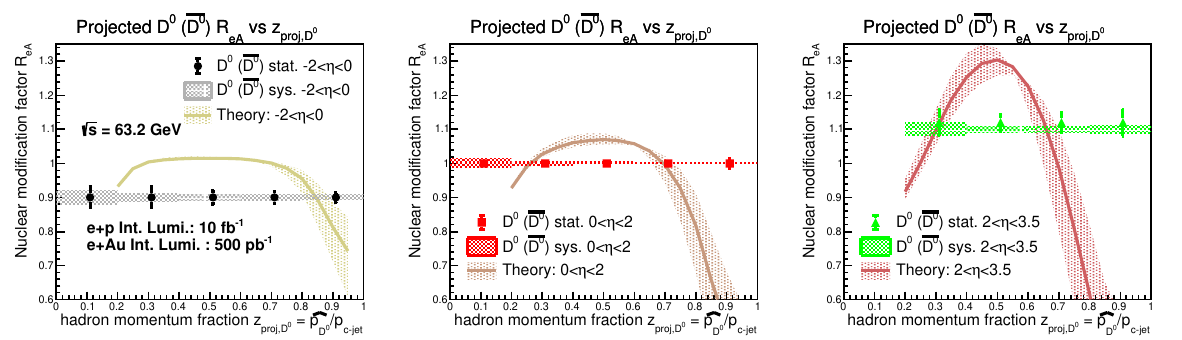}
\caption{Projected accuracy of hadron momentum fraction $z_{proj, D^{0}}$ dependent nuclear modification factor $R_{eAu}$ of reconstructed $D^{0}$ ($\bar{D^{0}}$) inside jet within $-2<\eta<0$ (left), $0<\eta<2$ (middle) and $2<\eta<3.5$ pseudorapidity regions in 63.2 GeV $e+p$ and $e+Au$ collisions, which are evaluated in simulation with 10~$fb^{-1}$ $e+p$ integrated luminosity, 500~$pb^{-1}$ $e+Au$ integrated luminosity, and the current ePIC detector performance.}
\label{fig-4}       
\end{figure}

The inclusive $\Lambda_{c}/D^{0}$ ratio in $e+p$ simulation reflects the string fragmentation functions implemented in PYTHIA, which is consistent with previous $e^{+}e^{-}$ and $e+p$ results. Though comparing the production of $D^{0}$ ($\bar{D^{0}}$) inside jet with the production of $\Lambda_{c}^{\pm}$ inside jet in $e+p$ and $e+A$ collisions within different kinematic phase space, more differential analyses will be carried out to map out the evolution of charm quark hadronization process under different nuclear medium conditions. Figure~\ref{fig-4} presents the projected accuracy of hadron momentum fraction $z_{proj, D^{0}}$ dependent nuclear modification factor $R_{eAu}$ of reconstructed $D^{0}$ ($\bar{D^{0}}$) inside jet within $-2<\eta<0$ (left), $0<\eta<2$ (middle) and $2<\eta<3.5$ pseudorapidity regions for 63.2~GeV $e+p$ and $e+Au$ collisions in comparison with latest Next-to-Leading Order theoretical calculations based on the parton energy loss model \cite{hf_th}. Future EIC heavy flavor hadron inside jet measurements will help extrapolating heavy quark fragmentation functions in vacuum and different cold nuclear media with great precision especially within the little constrained high hadron momentum fraction region.

\section{Summary and Outlook}
\label{sec-4}
The EIC project will enable a series of high precision heavy flavor hadron and jet measurements within a broad kinematic region under different nuclear medium conditions. Latest simulation studies have validated new physics capabilities such as exploring the universality of heavy quark fragmentation function in cold nuclear media utilizing the current conceptual design of the ePIC detector. Future EIC heavy flavor measurements will play an essential role in improving the current understanding of the final-state parton showering, heavy quark energy loss and hadronization mechanism within the little known kinematic region.
%
%
%

\end{document}